\documentclass[journal,letterpaper,10pt]{IEEEtran}
\usepackage{times}
\usepackage{flushend}
\usepackage{color}
\usepackage{epsf}
\usepackage{epsfig}
\usepackage{graphicx}
\usepackage{graphics}
\usepackage{epstopdf}
\usepackage[small]{caption}
\usepackage{amsmath}
\usepackage{amssymb}
\usepackage{amsthm}
\usepackage{amsxtra}
\usepackage[ruled]{algorithm2e}
\usepackage{algorithmic}
\usepackage{enumerate}
\usepackage{multirow}
\usepackage{enumerate}
\usepackage{amssymb}
\usepackage{lipsum}
\usepackage{setspace}
\usepackage{adjustbox}
\usepackage{multirow}
\usepackage{subcaption}
\usepackage{url}
\usepackage{color,soul}
\usepackage{geometry}
\usepackage[utf8]{inputenc}
\usepackage{array}
\usepackage{makecell}
\usepackage[normalem]{ulem}

\usepackage{rotating} % adde by Wayes Tushar
\usepackage{graphicx} % added by Wayes Tushar
\newcommand{\Rmnum}[1]{\expandafter\@slowromancap\romannumeral #1@}

\makeatother
\begin{document}
\title{Peer-to-Peer Trading in Electricity Networks: An Overview}
%\author{Wayes Tushar, Tapan K. Saha, Chau Yuen, David B. Smith, and H. Vincent Poor}
\author{Wayes~Tushar,~\IEEEmembership{Senior Member,~IEEE,}~Tapan~K.~Saha,~\IEEEmembership{Fellow,~IEEE,}~Chau~Yuen,~\IEEEmembership{Senior Member,~IEEE,}~David Smith,~\IEEEmembership{Member,~IEEE,}~ and~H.~Vincent~Poor,~\IEEEmembership{Fellow,~IEEE}
\thanks{W. Tushar and T. K. Saha are with the School of Information Technology and Electrical Engineering of the University of Queensland, Brisbane, QLD 4072, Australia (e-mail: wayes.tushar.t@ieee.org; saha@itee.uq.edu.au).}
\thanks{C. Yuen is with the Engineering Product Development Pillar of the Singapore University of Technology and Design (SUTD), 8 Somapah Road, Singapore 487372. (e-mail: yuenchau@sutd.edu.sg).}
\thanks{D. Smith is with the Data61, CSIRO, Australia. (e-mail: david.smith@data61.csiro.au)}
\thanks{H. V. Poor is with the Department of Electrical Engineering at Princeton University, Princeton, NJ 08544, USA. (e-mail: poor@princeton.edu).}
\thanks{This work is supported in part by the Queensland State Government under the Advance Queensland Research Fellowship AQRF11016-17RD2, in part by NSFC 61750110529, in part by NSoE\_DeST-SCI2019-0007, and in part by the U.S. National Science Foundation under Grant ECCS-1549881.}
}
\IEEEoverridecommandlockouts
\maketitle
%\thispagestyle{fancy}
%\thispagestyle{empty}
%\pagestyle{empty}
%\doublespace
% ================================
\begin{abstract}
Peer-to-peer trading is a next-generation energy management technique that economically benefits proactive consumers (prosumers) transacting their energy as goods and services. At the same time, peer-to-peer energy trading is also expected to help the grid by reducing peak demand, lowering reserve requirements, and curtailing network loss. However, large-scale deployment of peer-to-peer trading in electricity networks poses a number of challenges in modeling transactions in both the virtual and physical layers of the network. As such, this article provides a comprehensive review of the state-of-the-art in research on peer-to-peer energy trading techniques. By doing so, we provide an overview of the key features of peer-to-peer trading and its benefits of relevance to the grid and prosumers. Then, we systematically classify the existing research in terms of the challenges that the studies address in the virtual and the physical layers. We then further identify and discuss those technical approaches that have been extensively used to address the challenges in peer-to-peer transactions. Finally, the paper is concluded with potential future research directions.
\end{abstract}
% =================================
% \setcounter{page}{1}
 % ====================================
\section{Introduction}\label{sec:introduction} Over the last few years, there has been an extensive growth in small-scale distributed energy resources, which encompass behind-the-meter generation, energy storage, inverters, electric vehicles, and control loads. At the household level, in particular, the increase in the use of distributed energy resources has been unprecedented. For instance, the global market of rooftop solar photovoltaic (PV) panels is expected to grow by $11\%$ over the next six years, with an additional increase in residential storage systems from $95$ MW in $2016$ to $3700$ MW by $2025$~\cite{Peck_IEEESpectrum_Oct_2017}. These small-scale resources can be utilized not only to manage the energy demand more efficiently, but also to enable a significant mix of clean energy into the grid. However, to do so, it is important for the owners of these assets to act as proactive consumers --- referred to here as prosumers --- and actively participate in the energy market.

Given this context, feed-in-tariff (FiT) has been used extensively to enable prosumers to participate in energy trading~\cite{Cheng_AE_Oct_2017}. In FiT, prosumers with rooftop solar panels can sell their excess solar energy to the grid and can buy energy again from the grid in case of any energy deficiency~\cite{Tushar_Access_Oct_2018}. Unfortunately, the benefit to prosumers for participating in recent FiT schemes has been very marginal~\cite{Tushar_TIE_Apr_2015}. As a consequence, FiT schemes have been discontinued in some parts of the world such as the state of Queensland in Australia~\cite{Ref_5kW_QLD}.

As such, peer-to-peer (P2P) trading has emerged as a next generation energy management technique for smart grid that can enable prosumers to actively participate in the energy market either by selling their excess energy~\cite{Tushar_TSG_Press_2019} or by reducing the demand of energy via Negawatts, i.e., demand reduction or negative Watts~\cite{Jing_Negawatts_GTD_2019}. With the prosumers in control of setting the terms of transactions and the delivering of goods and services~\cite{Thomas_Nature_2018}, it is expected that the gain that the prosumers can reap from participating in P2P trading would be substantial~\cite{Tushar_AE_June_2019}. At the same time, the grid --- consisting of generators, retailers, and distribution network system provider (DNSPs) --- can also obtain significant benefit in terms of reducing peak demand~\cite{Zhang_AE_June_2018}, lowering investment and operational costs~\cite{Esther_AE_Jan_2018}, minimizing reserve requirements~\cite{Andoni_RSER_Feb_2019}, and improving power system reliability~\cite{Thomas_Nature_2018}.

However, trading in a P2P network is challenging. This is because, in P2P trading, it is expected that prosumers will trade their energy with one another with a very low (or, not any) influence from a central controller, which makes P2P platforms a trustless system. Hence, it is a challenging task to \emph{encourage prosumers} to cooperate in such a trustless environment~\cite{Tushar_Access_Oct_2018}. Further, in an energy system with a large number of users, it is difficult to \emph{model the decision making process} for various energy trading parameters given their rational choices that can conflict with the interests of other prosumers in the network~\cite{Tushar_SPM_July_2018}. Furthermore, electricity exchange is different from any other exchange of goods. This is due to the fact that prosumers are part of an electricity network, which has its own hard technical constraints on energy exchange\cite{Chapman_TSG_2018}. Completely decentralized P2P trading could be detrimental in maintaining the technical limit of the network within the safety range \cite{Chapman_TSG_2018}. Therefore, how to trade energy in the P2P network without compromising the network's security needs to be addressed. Finally, a number of stakeholders in the grid may request of prosumers P2P services with different objectives in mind. Thus, innovations are needed in the pricing scheme to prioritize these requests in order to deliver a non-congested service throughout the entire network, while keeping the network loss at a minimum~\cite{Baroche_TPWRS_July_2019}.

To that end, a large number of interesting results have been reported in the literature recently with the aim to address these challenges. Due to the complex nature of the problem as well as the broad range of techniques that have been used to solve it, to have a grasp of the entire paradigm of current P2P trading research has become a troublesome task. However, an overall understanding of the state-of-the art P2P research is important in order to: 1) initiate new research direction in this field; 2) cater for new challenges that are forthcoming in the energy sector; 3) develop more efficient and cost-effective energy trading mechanisms to deploy in real networks; and 4) design new services via P2P trading. Thus, having clear insight into the current state-of-the-art in P2P energy research could be beneficial for new researchers of power and energy systems. This is particularly true for investigators who want to contribute to developing a sustainable future through distributed energy resources.

Given this context, this paper aims to provide an overview of disruptive innovations in current P2P energy trading research contributing to revolutionizing the future energy sector by making the following contributions:
\begin{itemize}
\item We provide a background discussion of P2P networks, features of P2P trading, P2P energy markets, and an overview of challenges in P2P trading.
\item We determine the core technical approaches that are adopted by current studies to devise various solutions in P2P trading and provide a detailed discussion of each of the techniques.
\item We provide a number of potential research directions that would be valuable to investigate as extensions of current research practice.
\end{itemize}

We note that there are other recent review articles that discuss various aspects of P2P trading. For example, in \cite{Zhang_Procedia_May_2017}, the authors provide an overview of various P2P projects that are currently being implemented in different parts of the world. Extensive overviews of different types of P2P and community-based market frameworks are described in \cite{Sousa_RSER_Apr_2019} and \cite{Khorasany_IET_Dec_2018}. How different blockchain-based distributed ledger technologies can be applied for various applications in energy sector are reviewed in \cite{Wu_Sustainability_Aug_2018} and \cite{Troncia_Energies_Aug_2019}, whereas the challenges and opportunities of these applications are discussed in \cite{Andoni_RSER_Feb_2019}. Finally, optimization techniques and other technical approaches for energy trading are comprehensively reviewed in \cite{Abdella_Energies_June_2018}, and with a particular focus on game-theoretic application in \cite{Tushar_SPM_July_2018}.

Indeed, these existing review studies have contributed extensively to the body of energy trading knowledge that can provide researchers with a good understanding of various technical aspects of P2P trading. However, these reviews are suitable mainly for those who have some understanding of energy trading, P2P networks, and demand response management. Our paper, on the contrary, takes a step back and targets an audience with no or little prior knowledge of P2P trading and provides a basic understanding of most of the aspects of P2P trading including the definition, network elements, various layers, and market structure of P2P networks. Then, by introducing the challenges and solution approaches for different layers, this article helps readers to choose research directions in a specific layer and then have an in-depth understanding of the challenges and technical approaches relevant to that layer. Further, this paper is also different from existing studies in terms of organization and focus of discussion, which is mainly the trading approaches for addressing challenges of relevant layers. Note that this paper could also be a useful resource for experienced researchers for revising the understanding of the topic.

The rest of the paper is organized as follows. In Section~\ref{sec:Preliminaries}, we give an overview of the elements of P2P networks followed by a review of P2P market structures in Section~\ref{sec:market}. A detailed overview of state-of-the-art research in P2P trading is provided in Section~\ref{sec:OverviewofP2P} following a systematic classification. Key technical approaches that have been applied for P2P trading are identified and discussed in Section~\ref{sec:P2PTechnical}. Finally, Section~\ref{sec:conclusion} provides some concluding remarks with a list of potential future research directions.

\section{P2P Trading: Overview of Network Elements}\label{sec:Preliminaries}
\begin{figure}[t]
\centering
\includegraphics[width=\columnwidth]{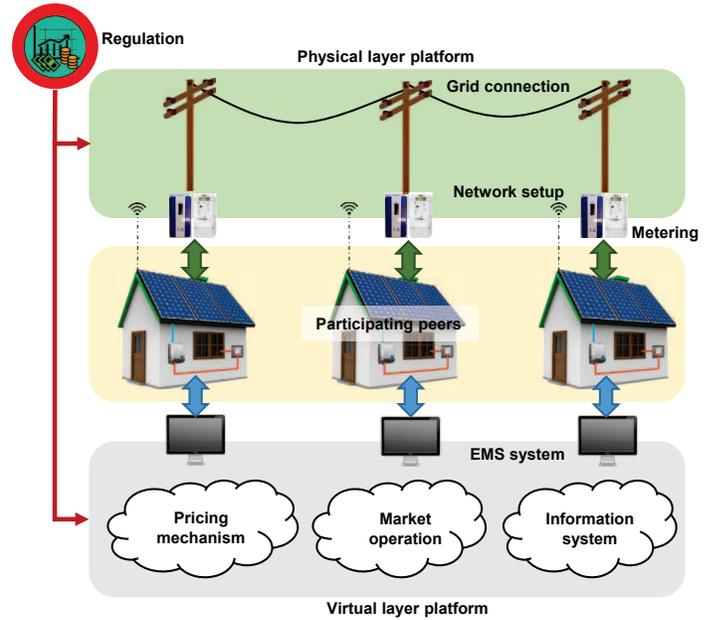}
\caption{An illustration of the physical layer and virtual layer platforms of a P2P energy network.}
\label{fig:P2PNetworkLayer}
\end{figure}
A distributed network architecture can be defined as a P2P network, if the participants of the network share a part of their own resources with one another. These shared resources provide the service and content offered by the network and can be accessed by other peers directly, without the intervention of intermediary entities~\cite{P2PDefinition_Aug_2001}. In addition, in a P2P network, any entity can be removed or added, if necessary, without the network suffering from any loss of network service. A formal definition of P2P networks can be found in \cite{Tushar_AE_June_2019}.

As shown in Fig.~\ref{fig:P2PNetworkLayer}, P2P network can be divided into two layers~\cite{Tushar_SPM_July_2018}: 1) virtual layer and 2) physical layer. The virtual layer essentially provides a secured connection for participants to decide on their energy trading parameters. It ensures that all participants have equal access to a virtual platform, in which transfer of all kinds of information takes place, buy and sell orders are created, an appropriate market mechanism is used to match the buy and sell orders, and finally, financial transactions are carried out upon successful matching of the orders.

The physical layer, on the other hand, is essentially a physical network that facilitates the transfer of electricity from sellers to  buyers once the financial settlements between both parties are completed over the virtual layer platform. This physical network could be the traditional distributed-grid network provided and maintained by the independent system operator or an additional, separate physical microgrid distribution grid, in conjunction with the traditional grid~\cite{Tushar_SPM_July_2018}. It has the necessary framework to enable communication between different prosumers and the grid. Here, it is important to note that the financial settlements between different prosumers in the virtual platform does not warrant  the physical delivery of electricity; rather, the payment can be thought of as indication from the buyers to their producing prosumers within the P2P network to process the injecting of renewable energy into the distribution grid.

Now, to successfully enable energy trading between different prosumers within the P2P network, it should have a number of key elements. A summary of these elements is given below. 
\begin{figure}[t]
\centering
\includegraphics[width=\columnwidth]{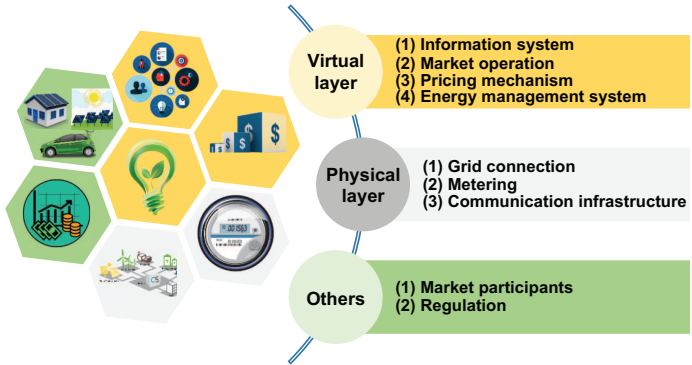}
\caption{Different elements of P2P network.}
\label{Fig:Elements}
\end{figure}
\subsection{Elements in the virtual layer}
\subsubsection{Information system}The heart of the P2P energy network is a high-performing and secured information system. The information system needs to be able to: 1) enable all market participants to communicate with one another for participating in energy trading; 2) integrate the participants within a suitable market platform; 3) give the participants equal access to the market; 4) monitor the market operation; and 5) set restrictions on participants' decisions to ensure network security and reliability. Examples of such information systems include blockchain-based smart contracts~\cite{Hou_TII_June_2019}, consortium blockchain~\cite{Li_TII_Aug_2018,Kang_TII_Dec_2017}, and Elecbay~\cite{Zhang_AE_June_2018}.
\subsubsection{Market operation}The information system of a P2P network facilitates the market operation consisting of market allocation, payment rules, and a clearly defined bidding format. The main objective of market operation is to enable the participants to experience an efficient energy trading process by matching the sell and buy orders in near real-time granularity. In market operations, energy generation of each producer influences the thresholds of a maximum and minimum allocation of energy. Different market-time horizons may exist in the market operation that should be able to produce enough allocation of energy at every stage of operation.
\subsubsection{Pricing mechanism}Pricing mechanisms are designed as parts of market operations and used to efficiently balance between the energy supply and demand. Pricing mechanisms used for P2P trading have a basic difference with that of the traditional electricity markets. For example, in traditional electricity markets, a significant portion of electricity price consists of electricity surcharges and taxes. However, as renewable energies typically have very low marginal costs~\cite{Esther_AE_Jan_2018}, prosumers can reap more profits by suitably setting prices for their energies. Nonetheless, pricing mechanisms need to reflect the state of energy within the P2P network, that is, a higher surplus of energy within the network should lower the energy price and vice versa.
\subsubsection{Energy management system}While participating in P2P trading via a particular bidding mechanism, the energy management system (EMS) of a prosumer secures its supply of energy. To that end, an EMS has access to the real-time supply and demand information of the prosumer through the transactive meter based on which it develops the generation and consumption profile of the prosumer and subsequently decides the bidding strategy to participate in the trading on behalf of the prosumer. For example, The EMS of a rational prosumer may always buy energy in the microgrid market when the price per unit of energy falls below its maximum price threshold~\cite{Tushar_SPM_July_2018}.
\subsection{Elements in the physical layer}
\subsubsection{Grid connection}P2P trading can be done for both grid-connected and islanded microgrid systems. For balancing the energy demand and generation in a grid-connected system, it is important to define the connection points of the main grid. By connecting smart meters at these connection points, it is possible to evaluate the performance of the P2P network, for example, in terms of energy and cost savings~\cite{Tushar_SPM_July_2018}. For islanded microgrids, on the other hand, participants should have enough generation capacity to ensure an appropriate level of security and reliability in supplying energy to consumers.
\subsubsection{Metering}Each prosumer should have appropriate metering infrastructure to be able to participate in P2P trading. In particular, each prosumer should be equipped with a transactive meter~\cite{Tushar_SPM_July_2018}, in addition to a traditional energy meter. A transactive energy is capable of deciding whether to participate in the P2P market based on the demand and generation data as well as the information available about market conditions (price, total demand, total available generation, and  network conditions). It can also communicate with other prosumers in the network by any appropriate communication protocol.
\subsubsection{Communication infrastructure}In P2P trading, the major requirement of communication is the discovery of prosumers and information exchange within the network. Multiple P2P communication architectures exist in the literature including structured, unstructured, and hybrid architectures~\cite{Jogunola_Energies_Dec_2018}. The choice of a communication architecture needs to fulfill the performance requirements recommended by the IEEE 1547.3-2007 for the integration of DER that include latency, throughput, reliability, and security~\cite{Jogunola_Energies_Dec_2018}.
\subsection{Other elements}
\subsubsection{Market participants}For P2P energy trading, the existence of a sufficient number of market participants within the network is necessary and a subgroup of the participants needs to have the capacity to produce energy. The purpose of P2P energy trading affects the design of pricing schemes and the market mechanism and therefore should be clearly defined. Further, the form of energy (that is, electricity or heat) traded.
\subsubsection{Regulation}The success of P2P trading in the future electricity market will probably be mostly governed by the regulation and energy policy. That is, governmental rules of a country decide what kind of market design will be allowed, how taxes and fees will be distributed, and how the P2P market will be into the existing energy market and supply systems. Thus, governments can support P2P energy markets to accelerate the efficient utilization of renewable energy resources and decrease environmental degeneration by regulatory changes. On the contrary, they can discourage the implementation of such markets as well if that impacts detrimentally current energy systems. An overview of the elements of P2P network is shown in \ref{Fig:Elements}.

\section{P2P Trading: Overview of Market Structure}\label{sec:market}
\begin{figure}[t]
\centering
\includegraphics[width=\columnwidth]{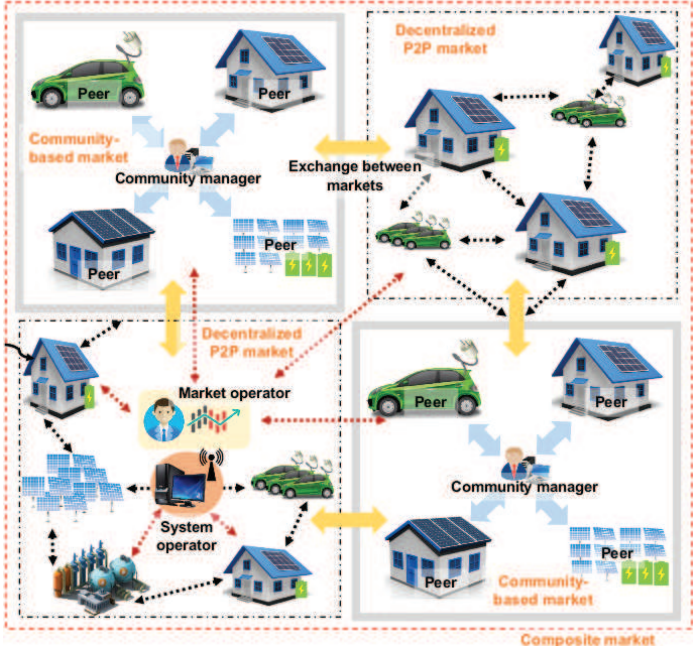}
\caption{Illustration of different types of markets as proposed in the literature for P2P trading.}
\label{fig:P2PMarket}
\end{figure}

In contrast to the top-down approach of the current energy market, P2P energy trading would require reorganizing electricity markets within decentralized management and collaborative principle that will allow for a bottom-up approach to empower prosumers~\cite{Sousa_RSER_Apr_2019}. Now, to determine how energy trading can be conducted in a P2P network, as shown in Fig.~\ref{fig:P2PMarket}, the market structures that have been proposed in the literature can be divided into three types: 1) Full decentralized markets; 2) Community-based markets; and 3) Composite markets.
\subsubsection{Fully decentralized market}In a fully decentralized P2P market, participating prosumers can independently and directly negotiate with one another to decide on the energy trading parameters without any centralized supervision. Such decentralization of a P2P market relies on bilateral contracts between individual prosumers as proposed in \cite{Thomas_TSG_Mar_2019}. Through the designed contract, \cite{Thomas_TSG_Mar_2019} captures both upstream-downstream energy balance and forward market uncertainty within the model. In \cite{Sorin_TPWRS_Mar_2019}, the authors propose another fully decentralized market for multi-bilateral economic dispatch, where prosumers with energy demand can choose their preferences for the type of energy source, such as local or green energy, for trading. Other examples of fully decentralized markets can be found in \cite{Esther_AE_Jan_2018} and \cite{Hug_TSG_July_2015}, where in \cite{Esther_AE_Jan_2018}, the authors discuss various properties of decentralized markets by referring to a test case of the Brooklyn microgrid. In \cite{Hug_TSG_July_2015}, on the other hand, the authors propose a distributed approach based on the
consensus and innovations method to coordinate local generation, flexible load, and storage devices within the microgrid to derive a distributed economic dispatch algorithm.
\subsubsection{Community-based market}A community-based P2P market can readily be applied to community microgrids~\cite{Paudel_TII_Aug_2019,Gonzalez_IEM_Dec_2018} and group of neighboring prosumers~\cite{ Tushar_TSG_May_2016}, in which the members of the community share common interests and goals even though they are not at the same location. The members may work either in a collaborative~\cite{Moret_TPWRS_EA_2018} or a competitive manner~\cite{Tushar_TSG_May_2016}. In a community-based P2P market, each member generally trades its energy within the community through a community manager. Indeed, a peer may also choose to trade its energy with someone outside the community, in which the community manager has a function associated with the energy exchanged with the outside world. Thus, a community manager manages the trading activities within the community, for example, by mimicking the role of an auctioneer~\cite{ Tushar_TSG_May_2016}, and also acts as an intermediary between the community and rest of the system~\cite{Sousa_RSER_Apr_2019}. Under community-based  P2P energy trading, the privacy of preferences and strategic schemes of each participant within the community are preserved~\cite{ Moret_TPWRS_EA_2018}. Further, the preferences of different classes of prosumers are reflected in their choice of energy parameters to trade within the community~\cite{Thomas_TPS_Early_2018}. A demonstration of a community-based market is shown in Fig.~\ref{fig:P2PMarket}.
\subsubsection{Composite market}A composite market is essentially a combination of fully decentralized and community-based markets in which each community and each single prosumer can interact with one another, while maintaining their own market properties. That is, on the one hand, each individual prosumer can engage in P2P trading between themselves, while also interacting with existing markets like fully distributed markets. On the other hand, a community manager can also oversee the trading inside a community. In such a market, prosumers may be nested into each other and form a community for trading within the neighborhood community. Examples of such markets can be found in \cite{Liu_TSG_Sept_2018} and \cite{Park_Sustainability_Feb_2018}.

Now, in a grid-connected system, a prosumer may need to deal with both regulated and deregulated P2P markets. Hence, how to integrate both of them in a single paradigm remains a challenge. Nonetheless, existing literature sheds some light on the possible ways for co-existence of such markets. For instance, in \cite{Tushar_TIE_Apr_2015}, the authors propose a three-party energy trading technique in which a community manager primarily participates in community-based P2P trading with the prosumers of its community in order to meet its demand for energy for maintaining different community facilities. However, the participation of the community manager in a regulated market for energy becomes necessary when it is unable to procure all required energy from the prosumers within a community. In that case, the community manager buys energy from a regulated market.

\begin{figure}[t]
\centering
\includegraphics[width=\columnwidth]{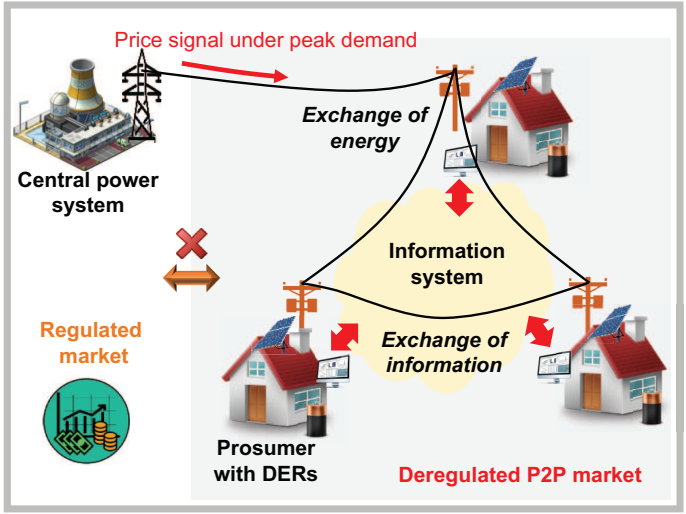}
\caption{Demonstration of how the coordination between regulated and deregulated P2P market with prosumers with distributed energy resources (DERs) is captured in \cite{Tushar_TSG_Press_2019}. Prosumers energy trading switches from regulated to deregulated P2P market with a price signal from the grid.}
\label{fig:MarketIntegration}
\end{figure}

Another interesting discussion of a potential integration scenario is proposed in \cite{Tushar_TSG_Press_2019}, in which prosumers primarily purchase their energy from a regulated electricity market as traditional customers. Now, when there is an extensive demand for energy from the grid, the grid sends a price signal to its selected prosumers to refrain from buying any energy from it for a specific period of time. Subsequently, prosumers form a fully decentralized P2P market among themselves and meet their demand for energy from their local generation. A graphical presentation of this market mechanism is shown in Fig.~\ref{fig:MarketIntegration}.

\section{P2P Trading: Overview of Existing Challenges}\label{sec:OverviewofP2P} Indeed, by participating in different energy market structures, on the one hand, the ultimate objective of P2P trading participants is to address a number of challenges related to energy trading including reducing the cost of energy usage, increasing and maintaining the sustainable use of renewable energy, and improving social engagement of prosumers. On the other hand, the decision making process of the prosumers to address these challenges are limited by the hard constraints imposed by the power network operators for ensuring the reliable operation of the power system without violating the voltage limit at prosumers' nodes, while keeping the overall network loss within reasonable limits. As such, what follows is an overview of how existing studies have developed P2P trading schemes as viable energy management techniques to solve various significant challenges of future smart grid.
\subsection{P2P Trading Challenges in Virtual Layer}\label{sec:P2PVirtualLayer}Studies proposed in the existing literature mainly focus on designing P2P energy trading mechanisms based on suitable pricing schemes that can enable participation of an extensive number of prosumers. Financial transactions are required to be securely conducted without involving a third party manager, while, at the same time, the trading should contribute to the achievement of desirable objectives of balancing local supply and demand, reducing prosumers' energy costs, and peak load shaving. As such, based on the focus of the study, the existing literature related to the virtual layer platform can be divided into five general categories as outlined bellow.
\begin{table*}[t]
\centering
\caption{Summary of different categories of studies that strive to achieve various objectives in the virtual layer and physical layer platforms.}
\small
\begin{tabular}{|m{2cm}|m{2.5cm}|m{9cm}|m{3cm}|}
\hline
\textbf{Different layers}&\textbf{Challenge} & \shortstack{\textbf{Overview of the study}} & \textbf{References}\\
\hline
\multirow{4}{*}{\thead{Virtual layer}}& Reducing cost of energy & To help reduce prosumers cost of energy by enabling small-scale prosumers with distributed energy resources to sell their excess energy to prosumers with energy deficiency. & \cite{Si_AE_Dec_2018,Hermana_ITSM_Fall_2016,Anees_AE_Nov_2019,Paudel_TII_Aug_2019,Luo_TPWRS_EA_2018,Al-Baz_AE_May_2019,Long_AE_Sep_2018,Wang_AE_Nov_2019,Nguyen_AE_Oct_2018,Meena_AE_Oct_2019,Tushar_AE_June_2019,Melendez_AE_Aug_2019,Long_AE_Sep_2018}\\
\cline{2-4}
&Balancing local generation and demand & To enable prosumers to coordinate their energy usage and prepare the buy and sell orders with the purpose of balancing the demand and supply within the community. & \cite{Tushar_Access_Oct_2018,Kang_TII_Dec_2017,Yang_TSMC_2019,Wang_TSMC_Aug_2019,Li_AE_Aug_2019,Chen_TSG_July_2019,Liu_TII_EA_2019,Paudel_TII_Aug_2019,Barbour_AE_Feb_2018}\\ \cline{2-4}
&Incentivizing \& engaging prosumers &To devise mechanism that will deliver prosumer-centric outcome to incentivize and engage prosumers to trade energy in the P2P network. & \cite{Kirchhoff_AE_June_2019,Tushar_AE_June_2019,Thomas_TPS_Early_2018,Tushar_Access_Oct_2018,Thomas_TSG_Mar_2019,Khorasany_TIE_EA_2019,Long_AE_Sep_2018,Luth_AE_Nov_2018,Saifuddin_Access_Apr_2019,Chen_TSG_July_2019,Kaixuan_AE_May_2019,Sorin_TPWRS_Mar_2019}\\\cline{2-4}
&Developing pricing mechanism &To design pricing mechanims that are suitable to apply for P2P nework for ensure fast and frequence trading. & \cite{Li_TII_Aug_2018,Thomas_TPS_Early_2018,Tushar_TSG_July_2017,Devine_TSG_Mar_2019,Cali_Access_June_2019}\\\cline{2-4}
&Identifying uncertainty and asynchronicity &To identify computation and communication complexity issues for robust system operation. &\cite{Moret_PSCC_June_2018,Moghaddam_IoT_Apr_2018, Anoh_ISGT_Oct_2018}\\\cline{2-4}
&Securing transactions &To enable prosumers to seamlessly engage in P2P trading through secure financial transactions among themselves.& \cite{Noor_AE_Oct_2018,Kang_TII_Dec_2017,Li_TII_Aug_2018,Wang_TSMC_Aug_2019,Zhang_AE_June_2018,Hou_TII_June_2019,Yang_TSMC_2019,Faizan_JEM_July_2019,Aitzhan_TDSC_Sept_2018,Naveed_IEM_2019}\\\hline
\multirow{2}{*}{\thead{Physical layer}}&Voltage \& capacity constraint&To prevent over voltage and reverse power flow issue due to P2P trading.& \cite{Imran_PESGM_Aug_2019,Espinosa_TPWRS_May_2016,Chapman_TSG_2018,Hamada_AE_Apr_2019,Hou_TII_June_2019,Williamson_JOE_2019}\\\cline{2-4}
&Network power loss &To understand the impact of P2P trading on network power loss and subsequent cost allocation between participants.& \cite{Nikolaidis_TPWRS_Early_2019,Baroche_TPWRS_July_2019,Xu_TIE_Nov_2019,Thomas_TPS_Early_2018}\\\cline{2-4}
&System strength &To understand the impact of increased use of renewable energy resources on the system strength of the power network.& \cite{Gu_CSEE_Sept_2019,Masood_TPWRS_Jan_2018,Masood_AE_Sept_2015,Wang_TPWRS_May_2018,Wu_TSE_July_2018,Zhou_TSTE_EA_2019}\\\hline
\end{tabular}
\label{table.ref1}
\end{table*}
\subsubsection{Reducing cost of energy}First category of studies propose how P2P trading can reduce prosumers cost of energy. Essentially, P2P trading enables small-scale prosumers with distributed energy resources to sell their excess energy, if any, to prosumers with energy deficiency, which has been shown to be very effective in reducing energy cost significantly~\cite{Si_AE_Dec_2018,Hermana_ITSM_Fall_2016}. Indeed, to facilitate such a trading mechanism, interaction among participating prosumers is the key~\cite{Paudel_TII_Aug_2019,Luo_TPWRS_EA_2018,Anees_AE_Nov_2019,Al-Baz_AE_May_2019}. The performance of P2P in terms of cost savings can further be improved if the batteries incorporated within the system also participate in the market~\cite{Long_AE_Sep_2018,Wang_AE_Nov_2019,Nguyen_AE_Oct_2018}. It is important to note that P2P trading is effective in reducing prosumers' costs for a number of energy trading scenarios including open market urban and remote systems~\cite{Meena_AE_Oct_2019}, fully decentralized systems~\cite{Tushar_AE_June_2019,Melendez_AE_Aug_2019}, and community-based microgrids~\cite{Long_AE_Sep_2018}.
\subsubsection{Balancing local generation and demand}The reduction in energy cost in P2P trading is due to its ability to enable prosumers with deficiency to meet their demand by buying the required energy from prosumers with energy surplus at a cheaper rate~\cite{Tushar_TIE_Apr_2015,Tushar_Access_Oct_2018} compared to the traditional market. However, such trading in a local environment necessitates the balance of supply and demand of energy within the community, which is the focus of the second category of studies. Now, for balancing the demand and supply, a ledger is necessary that can track all the transactions as well as the available supply from and demand of each participating prosumers. In P2P trading, this is currently achieved by using blockchain based platforms, as shown in~\cite{Kang_TII_Dec_2017,Yang_TSMC_2019,Wang_TSMC_Aug_2019} and \cite{Li_AE_Aug_2019}. Under the blockchain platform, prosumers learn the energy usage pattern of different sellers and buyers~\cite{Chen_TSG_July_2019}, manage their own energy consumption through residential demand response schemes~\cite{Liu_TII_EA_2019}, and then trade with one another, whenever applicable, within the local community~\cite{Paudel_TII_Aug_2019} to keep the balance between supply and demand. Indeed, if there is still an imbalance between the supply and demand, that can be fulfilled by the grid~\cite{Tushar_Access_Oct_2018}, community storage~\cite{Barbour_AE_Feb_2018}, or diesel generator~\cite{Zhang_TIA_July_2016} in expense of higher costs.
\subsubsection{Incentivizing \& engaging prosumers}Clearly, to successfully reap the benefits explaining in the previous two sections, prosumers need to be actively involved in the trading mechanism~\cite{Kirchhoff_AE_June_2019}. This is only possible if prosumers find outcomes of the P2P trading beneficial for them. Hence, the mechanisms need to be prosumer-centric~\cite{Tushar_AE_June_2019} (also, known as consumer-centric~\cite{Tushar_TSG_May_2014}). The third category of existing literature devote its efforts to determine how to incentivize prosumers to extensively participate in P2P trading. Under this category, a large number of techniques have been proposed to ensure prosumer-centric delivery of outcomes including multi-class energy management~\cite{Thomas_TPS_Early_2018}, motivational psychology~\cite{Tushar_Access_Oct_2018}, bilateral contract theory~\cite{Thomas_TSG_Mar_2019,Khorasany_TIE_EA_2019}, reinforcement learning~\cite{Saifuddin_Access_Apr_2019,Chen_TSG_July_2019}, game theory~\cite{Tushar_Access_Oct_2018}, prediction-integrated double auction~\cite{Kaixuan_AE_May_2019}, consensus-based approach~\cite{Sorin_TPWRS_Mar_2019}, and aggregated battery control technique~\cite{Long_AE_Sep_2018,Luth_AE_Nov_2018}.
\subsubsection{Developing pricing mechanism}Extensive prosumers' engagement in P2P trading and the subsequent benefits exclusively depend on the financing transactions among the participating buyers and sellers in the trading. Therefore, there is a need for innovative pricing schemes that are particularly applicable for P2P trading, which is the main aspect of the fourth category of study in the existing literature. For example, in \cite{Li_TII_Aug_2018}, a credit-based pricing scheme is proposed for fast and frequent energy trading. Based on different type of prosumer classes, a distributed price-directed optimization scheme is studied in \cite{Thomas_TPS_Early_2018}. A discrimination pricing scheme suitable to deploy in P2P network is discusses in \cite{Tushar_TSG_July_2017}. Other example of different pricing schemes can also be found in \cite{Devine_TSG_Mar_2019} and \cite{Cali_Access_June_2019}.
\subsubsection{Identifying uncertainty and asynchronicity}While P2P markets have substantial advantages in terms of product differentiation, customer involvement, and low transaction costs, the market outcome could be suboptimal if interaction and negotiation mechanisms are not adequately designed. In particular, when a large number of prosumers engage in P2P transactions, computation and communication complexity issues must be resolved for robust system operation. As such, extensive computational analysis of existing decentralized and distributed algorithms is provided in \cite{Moret_PSCC_June_2018}. It is identified that both computation and communication complexity impact the average time per iteration. Computation delays appear in cases of non-performing hardware and solving complicated optimization sub-problems. Communication delays are caused by bandwidth limits or internet traffic. Other examples of such studies can be found in \cite{Moghaddam_IoT_Apr_2018} and \cite{Anoh_ISGT_Oct_2018}.
\subsubsection{Securing transactions}To seamlessly engage prosumers to perform P2P trading, the financial transactions among themselves need to carry out securely. Further, the buy and sell orders and price information also need to be available over a secured platform for the prosumers to be incentivized and participate in the trading~\cite{Noor_AE_Oct_2018}. As such, the fifth and final category of studies, as outlined in this paper, deal with securing P2P transactions over the virtual layer platform. Different types of secured transactions are modeled based on the blockchain platform. For example, consortium blockchains are implemented in \cite{Kang_TII_Dec_2017} and \cite{Li_TII_Aug_2018} for conducting P2P trading for electric vehicles and Industrial Internet-of-Things respectively. For energy trading in a residential community, some popular examples of secured transaction platform include Hyperledger~\cite{Wang_TSMC_Aug_2019}, Elecbay~\cite{Zhang_AE_June_2018}, smart contracts~\cite{Hou_TII_June_2019,Yang_TSMC_2019}, Ethereum~\cite{Faizan_JEM_July_2019}, and multi-signature blockchain~\cite{Aitzhan_TDSC_Sept_2018}.

We note that other than the discussed categories, P2P trading is also conducted to achieve peak load shaving~\cite{Wang_AE_Oct_2019} and setup virtual power plant~\cite{Thomas_Nature_2018} to provide ancillary services to the grid. A summary of the existing studies that deal with P2P trading in the virtual layer platform is given in Table~\ref{table.ref1}.
\begin{figure*}[t]
\centering
\includegraphics[width=0.65\linewidth]{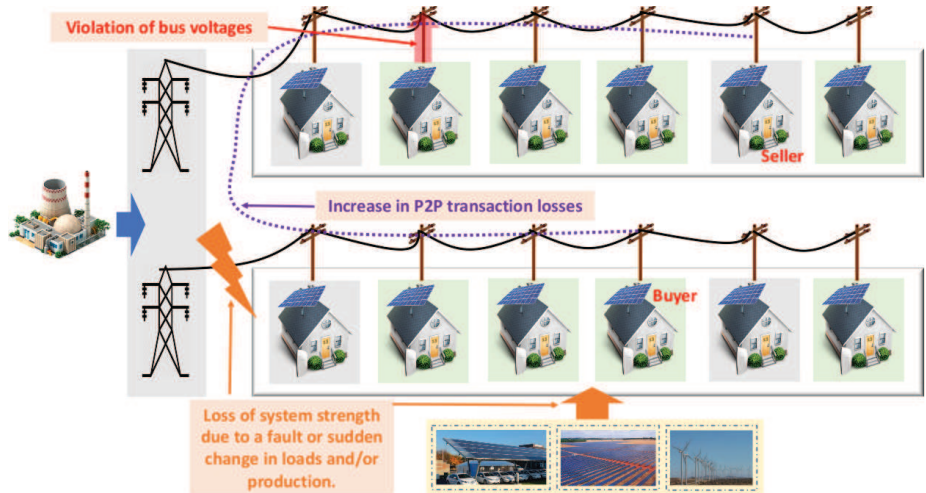}
\caption{Illustration of challenges of the physical layer of P2P network.}
\label{Fig:PhysicalLayerChallenges}
\end{figure*}
\subsection{P2P Trading Challenges in Physical Layer}\label{sec:P2PPhysicalLayer}
The state-of-the-research in the virtual layer platform has laid the foundation for performing P2P trading in the energy network by capturing decision making process of prosumers in secured and transparent energy trading platforms as well as devising pricing scheme to ensure their extensive participation. However, once the decision of the energy trading parameters is established at the virtual layer, the actual transfer of an agreed amount of energy is transferred over the physical layer. Now, the power system puts hard constraints on the exchange of energy over its network~\cite{Chapman_TSG_2018}. As a consequence, if the decision making process in the virtual layer platform does not consider the potential impact of P2P trading of energy on the physical layer platform, the transfer of energy may violate a number of technical constraints. For example, in \cite{Imran_PESGM_Aug_2019}, we discuss the feasibility of P2P trading in a grid-connected power network, in which it is shown that some of the bus voltages may exceed the network imposed voltage limit and compromise the security and reliability of the network if P2P trading is coordinated ignoring the constraints of the network.

Given this context, recently, there has been a growing interest to address challenges that may impede the transfer of energy over the physical layer platform of P2P networks. In particular, three types of network challenges have been being studied in the existing literature: 1) Violation of voltage and capacity constraints, 2) increase in network power loss, and 3) loss of system strength. An overview of these challenges is illustrated in Fig.~\ref{Fig:PhysicalLayerChallenges}.

\subsubsection{Violation of voltage \& capacity constraints}Since, residential users who act as prosumers are connected to low voltage distribution systems, their active participation in P2P trading could cause an over voltage issue~\cite{Imran_PESGM_Aug_2019} and reverse power flow~\cite{Espinosa_TPWRS_May_2016}. To combat such cases, a novel methodology based on sensitivity analysis to assess the impact of P2P transmission on the network and the subsequent cost with the associate energy exchange is studied in \cite{Chapman_TSG_2018}. Now, the inverters installed within the prosumers' premises primarily responsible for pushing power to the network and extensive transaction of energy over the P2P network would obviously increase the load on the inverters. This can be reduced by exploiting grid voltage support algorithms for smart photovoltaic inverters, based on distributed optimization and peer-to-peer communication~\cite{Hamada_AE_Apr_2019}. Another resulting detrimental impact of pushing power by an extensive number of prosumers in the network could be the operational overhead, which can increase the expense of energy transportation due to the requirement of a long chain maintaining of many blocks~\cite{Hou_TII_June_2019}. A blockchain-based P2P trading scheme exploiting local energy storage is shown to be effective to avoid such scenarios in \cite{Hou_TII_June_2019}. Finally, how large prosumsers like community-microgrids that are operated in multiple voltage levels can participate in energy transfer is discussed in \cite{Williamson_JOE_2019}.
\subsubsection{Increase in network power loss}While power exchange between sellers and buyers via P2P trading could increase the node voltages and overload the network capacity, it inevitably incurs losses as well. Consequently, this entails extra energy amounts and costs, above the local net demand, that needs to be produced and recovered by each market entry~\cite{Nikolaidis_TPWRS_Early_2019}. To that end, a graph-based loss allocation scheme to harmonize the physical attributes of the low voltage distribution grid is proposed and tested in \cite{Nikolaidis_TPWRS_Early_2019}. Another cost allocation mechanism is proposed in \cite{Baroche_TPWRS_July_2019}, in which costly incentives are used to allocate P2P market related grid costs to the participants by the system operator. In this allocation process, a degree of freedom is given to the system operator to reach cost recovery. Other than cost allocation, another interesting mechanism to reduce the loss during P2P trading could be to choose an optimal power routing strategy \cite{Xu_TIE_Nov_2019}. By doing to, as shown in \cite{Xu_TIE_Nov_2019}, it is possible to optimize power dispatching with the minimum power loss ratio between the power seller and buyer within the P2P network. Finally, energy classes are introduced in \cite{Thomas_TPS_Early_2018} to allow energy to be treated as a heterogeneous product and to coordinate P2P energy trading to minimize the cost associated with network losses.
\subsubsection{Loss of system strength}Synchronous generators make major contributions to re-stabilize power systems following voltage/frequency disturbances by providing system strength and inertia~\cite{Gu_CSEE_Sept_2019}. However, as technologies like P2P trading penetrate the market, the growth of renewable energy sources within the system has been extensive. This necessitates the retirement of a large number of synchronous generators in recent times. As a consequence, maintaining system strength in renewable dominated power networks is becoming more and more challenging~\cite{Masood_TPWRS_Jan_2018}. A typical example of system failure due to lack of system inertia can be found in the recent blackout in South Australia~\cite{Masood_AE_Sept_2015}. 

As such, the necessity of investigating the impact of renewable energy sources on the system strength has become very important in the current context. Some example of such studies can be found in \cite{Wang_TPWRS_May_2018,Wu_TSE_July_2018} and \cite{Zhou_TSTE_EA_2019}. In \cite{Wang_TPWRS_May_2018}, a real-time method is designed for solar plants that coordinates photovoltaic inverters and battery storage systems in order to provide voltage regulation. The authors in \cite{Wu_TSE_July_2018} propose a site-dependent short-circuit ratio to analyze the system strength and voltage variability of renewable dominated power systems. Finally, \cite{Zhou_TSTE_EA_2019} considers the characteristics of diesel engines such as time-delay management, spinning reserve strategies, and ramp rate to develop an optimal photovoltaic-storage control strategy  and capacity in order to combat the variability of photovoltaic output.

Now, clearly, implementing P2P trading in power system networks is challenging. As such, to deliver trading schemes that address the challenges of both virtual layer and physical layer platforms simultaneously, a number of technical approaches have been adopted. To this end, what follows in the next section is an overview of technical approaches that have been proven to be effective in modeling P2P trading in the virtual and physical layer platforms.
\section{P2P Trading: Overview of Technical Approaches}\label{sec:P2PTechnical}Based on the approaches adopted by recent studies, four general techniques can be identified as the main contributors to the design of existing P2P energy trading schemes. They are 1) game theory, 2) auction theory, 3) constrained optimization, and 4) blockchain. An overview of different technical approaches adopted by existing studies is summarized in Table~\ref{table:summary}.
\begin{table*}[t]
\centering
\caption{Summary of the different technical approaches adopted to enable P2P energy trading.}
\small
\begin{tabular}{|m{2cm}|m{7cm}|m{4cm}|m{2cm}|m{2cm}|}
\hline
\textbf{Technical approach}&\textbf{General focus of the approach} & \textbf{Popular method} &\textbf{Literature in virtual layer} & \textbf{Literature in physical layer}\\
\hline
\textbf{Game theory}& To capture the competition and cooperation between different participants of P2P energy trading market to deliver a solution that is stable, sometime optimal, and mutually beneficial for all involved parties& Stackelberg game, coalition formation game, canonical coalition game, non-cooperative Nash game, generalized Nash game. & \cite{Tushar_TSG_Press_2019,Paudel_TII_Aug_2019,Anees_AE_Nov_2019,Wang_AE_Nov_2019,Li_TII_Aug_2018,Melendez_AE_Aug_2019,Yang_TSMC_2019,Zhang_AE_June_2018,Li_AE_Aug_2019,Thomas_TSG_Mar_2019,Bhatti_AE_Nov_2019,Noor_AE_Oct_2018,Wang_AE_Oct_2019,Luo_TPWRS_EA_2018,Lee_JSAC_July_2014,Tushar_Access_Oct_2018}&\cite{Zhang_AE_Feb_2019}\\
\cline{1-5}
\textbf{Auction theory}&To capture the interaction between a number of sellers and buyers of a P2P market so as to enable them to trade their electricity in a step-by-step fashion.& Double auction& \cite{Tushar_TSG_Press_2019,Kang_TII_Dec_2017,Liu_TII_EA_2019,Wang_AE_Oct_2019,Kaixuan_AE_May_2019} & \cite{Chapman_TSG_2018}\\ \cline{1-5}
\textbf{Constrained optimization}&To use mathematical programming technique for optimizing the parameters of P2P trading under different hard and soft constraints imposed by the market and power system. &LP, MILP, ADMM, NLP& \cite{Luth_AE_Nov_2018,Nguyen_AE_Oct_2018,Thomas_TPS_Early_2018,Long_AE_Sep_2018,Si_AE_Dec_2018,Khorasany_TIE_EA_2019}&\cite{Baroche_TPWRS_July_2019,Hamada_AE_Apr_2019,Xu_TIE_Nov_2019}\\\cline{1-5}
\textbf{Blockchain}&To provide a data structure that can be replicated and shared among members to enable secured, transparent, and decentralized energy trading in a P2P network.&Smart contract, Elecbay, consortium blockchain, Hyperledger, Ethereum& \cite{LeeThomas_Nature_2019,Luo_TPWRS_EA_2018,Kang_TII_Dec_2017,Li_TII_Aug_2018,Aitzhan_TDSC_Sept_2018,Yang_TSMC_2019,Li_AE_Aug_2019,Wang_TSMC_Aug_2019,Zhang_AE_June_2018,Devine_TSG_Mar_2019,Noor_AE_Oct_2018,Faizan_JEM_July_2019}& Not available\\\cline{1-5}
\end{tabular}
\label{table:summary}
\end{table*}
\subsection{Game theory}
\subsubsection{Preliminary}Game theory is a mathematical tool that analyzes the strategic decision making process of a number of players in a competitive situation, in which the decision of action taken by one player depends on and affects the actions of other players~\cite{Basar_Book_1995}. Game theory can generally be divided into two categories: non-cooperative games and cooperative games.

\emph{Non-cooperative game:}~In non-cooperative games, the strategic decision-making process of a number of independent players that have partially or completely conflicting interests are analyzed to determine outcomes that are influenced by their actions. In such games, players take their decisions without communicating with one another. Any cooperation that may arise in a non-cooperative game cannot be a result of either communication or coordination of strategic choices among players~\cite{Saad_SPM_Sept_2012}.

In general, two types of non-cooperative games have been used for designing energy trading schemes: static games and dynamic games. In a static game, players take their actions once only, either simultaneously or at different times. In a dynamic game, on the other hand, time plays a central role in the decision making process of each player. Players in a dynamic game act more than once and have inputs regarding the choices of other players.

The most popular solution concept of a non-cooperative game is the Nash equilibrium~\cite{Tushar_TSG_Dec_2012}. Essentially, a Nash equilibrium refers to a stable state of a non-cooperative game in which no player can be better paid off by unilaterally deviating from its action, provided all other players are also taking their Nash equilibrium strategies. For instance, let a static game $\Gamma$ be defined by $\Gamma = \{\mathcal{N}, \mathbf{s}_n, U_n\}$, where $\mathcal{N}$ is the set of all players participating in the game, $s_n$ is the vector of strategies of player $n\in\mathcal{N}$, and $U_n$ is the utility function of $n$ that reflects the benefit that the player $n$ can reap by choosing a strategy $s_n$. Now, the Nash equilibrium of $\Gamma$ can be defined as $\{\mathbf{s}^*: \mathbf{s}^* = [s_n^*, \mathbf{s}_{-n}^*], U_n(\mathbf{s}^*)\geq U_n(s_{n}, \mathbf{s}_{-n}^*)\}$. Here, $\mathbf{s}_{-n}^*$ is the strategy vector of players in $\mathcal{N}\setminus \{n\}$.

A particular non-cooperative game that has extensively been used to design P2P trading in the literature is the Stackelberg game~\cite{Tushar_TSG_Dec_2012}. A Stacklelberg game is essentially a strategic game in which at least one player is defined as the leader who makes its decision first and commit a strategy before other players. Other players, on the other hand, act as the followers in the game, who optimize their strategies in response to the action taken by the leader. The solution concept of a Stackelberg game is the Stackelberg equilibrium, in which followers participate in a non-cooperative Nash game among themselves and reach a Nash equilibrium in response to the Leader's decision. At the Stackelberg equilibrium, neither the leader nor any follower has any incentive to deviate from its chosen strategy~\cite{Tushar_TSG_May_2016,Tushar_TSG_Press_2019}.

\emph{Cooperative game:}~Cooperative games, also known as coalitional games, deal with incentives that can make independent decision makers to act together as one entity to improve their position in the game. The most common form of coalitional game is the characteristic form~\cite{Saad_SPM_Sept_2012}, where the value of coalition is determined by the members of the coalition irrespective of the structure of the coalition. Now, coalition games can be classified into three types.
\paragraph{Canonical coalition game}~In canonical coalition games, forming a grand coalition with all players is never detrimental to any participant of the game. Consequently, the main objectives of such a game is to determine whether or not a grand coalition can be formed, to investigate if the grand coalition is stable, and to formulate a fair revenue distribution scheme for distributing the gains of coalition among the players. The most commonly considered solution concept of a canonical coalition game is the core~\cite{Saad_SPM_Sept_2012}. Meanwhile, for revenue distribution, the most popular methods include the Shapley value, the Kernel, the nucleolus, and the strong epsilon-core.
\paragraph{Coalition formation game}~The objective of static coalition formation game is to study the network coalitional structure. In a dynamic coalitional game, on contrary, the game is subject to environmental changes, including a change in the number of players or a variation in network topology. Therefore, the main objective of this type of dynamic game is to study the formation of a coalitional structure through players' interactions and inquire the properties of the structure and its adaptability to environmental variations~\cite{Tushar_SPM_July_2018}.
\paragraph{Coalitional graph game}~Coalition graph games deal with the connectivity of communications between players of the game. The main objectives are to derive low complexity distributed algorithms for players who want to build network graphs and to study the properties of the graphs~\cite{Saad_SPM_Sept_2012}.
\subsubsection{Game theory for P2P trading in the virtual layer}\label{sec:P2PVirtualLayer}In the virtual layer platform, game theory has been extensively used to obtain different objectives that are outlined in Section~\ref{sec:P2PVirtualLayer}. For example, Stackelberg game has been used to reduce the cost of energy~\cite{Paudel_TII_Aug_2019,Anees_AE_Nov_2019,Wang_AE_Nov_2019} and to design suitable pricing scheme for secured transaction in P2P trading~\cite{Li_TII_Aug_2018}. The application of non-cooperative Nash games in P2P trading can be found for reducing energy cost~\cite{Melendez_AE_Aug_2019}, balancing local generation and demand~\cite{Yang_TSMC_2019,Zhang_AE_June_2018,Li_AE_Aug_2019}, encouraging prosumers' participation in the trading~\cite{Thomas_TSG_Mar_2019,Bhatti_AE_Nov_2019}, improving the security of transaction~\cite{Noor_AE_Oct_2018} and peak shaving~\cite{Wang_AE_Oct_2019}. Finally, the authors in \cite{Luo_TPWRS_EA_2018}, \cite{Lee_JSAC_July_2014}, \cite{Tushar_Access_Oct_2018}, and \cite{Tushar_AE_June_2019} demonstrate how the framework of a canonical coalition game can be used to obtain reduction in energy cost via balancing local generation and demand, fairness in deciding the trading price, and increased participation of prosumers in P2P trading respectively.

\subsubsection{Game theory for P2P trading in the physical layer}\label{sec:P2PPhysicalLayer}In the physical layer platform, however, the application of game theory has been limited so far. One application of a multiple-leader-multiple-follower Stackelberg game can be found in \cite{Zhang_AE_Feb_2019} with an objective to study the influence level of transmission losses on trading behavior of retailers and consumers. In particular, the authors propose a credit rating based optimal pricing and energy scheduling model by considering retailers as leaders and consumers as followers. It is shown that transmission losses cannot be ignored in energy trading, which, if avoided, could result in a large difference between actual power received by consumers and their demands.
\subsection{Double Auction}
\subsubsection{Preliminary}A double auction involves a market of a number of buyers and sellers seeking to interact so as to trade their goods~\cite{Saad_SmartGridComm_2011}. In a double auction, potential buyers submit their bids to an auctioneer while, at the same time, potential sellers simultaneously ask prices to the auctioneer. This is usually done through a step-by-step process as follows~\cite{Huang_CI_Nov_2002}:
\begin{enumerate}[(1)]
\item Sellers submit their reservation prices in an increasing order.
\item Buyers are arranged in a decreasing order of their reservation bids.
\item Once the sellers and buyers orders are ordered, the aggregated supply and demand curves are generated that meet at a intersection point.
\item The intersection point establishes the auction price and the number of seller and buyers that eventually engage in the market trading process.
\end{enumerate}

In the double auction process, the sellers and buyers need to truthfully report their reservation prices and bids for efficient operation of the market. Hence, auction mechanisms need to satisfy the properties of individual rationality and incentive compatibility~\cite{Tushar_TSG_May_2016}. Now, a double auction scheme is said to possess the property of \emph{individual rationality} if the utility that a prosumer receives for participating in the auction mechanism cannot be improved otherwise, provided all other prosumers in the auction are choosing their selected strategies. Meanwhile, a double auction mechanism is called incentive-compatible, if every participant of the auction mechanism can achieve the best outcome to themselves by acting upon their true preference that they revealed during the above mentioned Steps 1 and 2.
\subsubsection{Double auction for P2P trading in the virtual layer}In the virtual layer platform, double auction technique has been used by \cite{Kang_TII_Dec_2017,Liu_TII_EA_2019,Wang_AE_Oct_2019,Kaixuan_AE_May_2019} to achieve objectives of  balancing local generation and demand, shaping the demand at the peak hour, and improving prosumers engagement in the trading. For example, the authors in \cite{Kang_TII_Dec_2017} propose a consortium blockchain enabled double auction mechanism to decide on the electricity price and traded energy amount by the prosumers in order to maintain the balance between the local generation and supply. For a similar purpose, an optimal bidding strategy via a double auction is proposed in \cite{Liu_TII_EA_2019} for residential houses. In \cite{Wang_AE_Oct_2019} and \cite{Kaixuan_AE_May_2019}, the authors exploit a Nash bargaining model and data-driven prediction-integration models respectively to design the double auction frameworks with the purpose of reducing peak demand and improving prosumers' engagement in P2P trading.
\subsubsection{Double auction for P2P trading in the physical layer}The application of double auction to address problems related to physical layer platform is elaborated in \cite{Chapman_TSG_2018}. In particular, \cite{Chapman_TSG_2018} propose a decentralized P2P architecture that can facilitate local energy trading. In designing the scheme, the authors explicitly take into account the underlying network constraints at the distribution level. The market mechanism is developed using a continuous double auction technique, a simple market format that matches parties interested in trading, rather than holding any of the traded commodity itself. As such, this scheme is very well suited for P2P exchanges. It is further shown in \cite{Chapman_TSG_2018} that in continuous double auctions comprising bidders with rational goals (i.e. participants only trade at a profit), trades are always Pareto-improving. Subsequently, the continuous double auction moves towards an allocation that is Pareto efficient with a highly efficient allocation of commodities \cite{good_JPE_Feb_1993}.
\subsection{Constrained optimization}
\subsubsection{Preliminaries}A number of constrained optimization techniques have been used to design P2P energy trading schemes. Examples of some techniques include linear programming (LP), mixed integer linear programming (MILP), alternating direction method of multipliers (ADMM), and nonlinear programming (NLP).
\paragraph{LP}LP is a mathematical programming technique to achieve the optimal outcome in a mathematical model, in which all requirements are presented by linear forms. Any LP can be expressed in its canonical form as
\begin{eqnarray}
\text{Maximize}~\mathbf{b}^T\mathbf{x},\label{Eqn:Objfunc1}
\end{eqnarray}
$\text{such that}~\mathbf{A}\mathbf{x}\leq\mathbf{c},~\text{and}~\mathbf{x}\geq 0$. Here, $\mathbf{x}$ is the vector of variables to be determined, $\mathbf{c}$ and $\mathbf{b}$ are coefficient vectors, $\mathbf{A}$
 is a matrix of coefficients, and ${\displaystyle (\cdot )^{\mathrm {T} }}$ is the transpose of  $(\cdot )$. The expression in \eqref{Eqn:Objfunc1} is called the objective function and the inequalities $\mathbf{Ax} \leq \mathbf{c}$ and $\mathbf{x} \geq 0$ are known as the constraints that need to be satisfied.
\paragraph{MILP}MILP is a special case of integer linear programming, in which only some of the variables are constrained to be integers and, unlike integer linear programming, other variables are allowed to be non-integer. Mathematically, an MILP can be expressed as same as an integer linear programming
\begin{eqnarray}
\text{Maximize}~\mathbf{b}^T\mathbf{x},\label{Eqn:Objfunc2}
\end{eqnarray}
$\text{subject to}~\mathbf{A}\mathbf{x}+ \mathbf{s}=\mathbf{c},~\mathbf{x}\geq 0$, $\mathbf{s}\geq 0$, and $\mathbf{x}\in\mathbb{Z}^n$, where some of the entries are not integer.
\paragraph{ADMM}ADMM is an algorithm that solves convex optimization problems by breaking them into smaller pieces, each of which are then easier to handle~\cite{Boyd_Book_2011}. Essentially, ADMM is a variant of augmented Lagrangian scheme that uses partial updates for the dual variables. This can be mathematically expressed for a maximization problem  as
\begin{eqnarray}
{\displaystyle\text{Maximize}_{\mathbf{x,z}}}f(\mathbf{x}) + g(\mathbf{z}),\label{Eqn:Objfunc3}
\end{eqnarray}
subject to $\mathbf{Ax} + \mathbf{Bz} = \mathbf{c}$, where is $\mathbf{z}$ is a vector of second variables. Thus, ADMM can have two objectives with two separate sets of variables.
\paragraph{NLP}NLP is a mathematical technique that solves an optimization problem, in which the objective function is non-linear and/or the feasible region is determined by nonlinear constraints. In maximization form, the problem can be expresses as that in \eqref{Eqn:Objfunc1} with nonlinear objective function and/or nonlinear constraints.

\subsubsection{Constrained optimization for P2P trading in the virtual layer}Different constrained optimization techniques have been heavily utilized in the literature to design P2P energy trading techniques in the virtual layer platform. For example, in \cite{Luth_AE_Nov_2018}, the authors exploit a LP approach to design a novel multi-energy management strategy based on the complementarity of multi-energy demand with the purpose to explore optimal energy scheduling problems of prosumers. An MILP technique is used by \cite{Nguyen_AE_Oct_2018} to optimize the use of energy generated from the rooftop solar with battery for P2P energy trading. A multi-class energy management technique with the purpose of P2P trading is designed in \cite{Thomas_TPS_Early_2018} using ADMM optimization technique. The application of NLP in \cite{Long_AE_Sep_2018} is for designing a P2P energy sharing mechanism in a community through aggregated battery control. Further use of constrained optimization in P2P trading can be found in \cite{Si_AE_Dec_2018} and \cite{Khorasany_TIE_EA_2019}.
\subsubsection{Constrained optimization for P2P trading in the physical layer} In the physical layer, the most popular constrained optimization technique has been ADMM as can be seen from its application in \cite{Baroche_TPWRS_July_2019} and \cite{Hamada_AE_Apr_2019}. In \cite{Baroche_TPWRS_July_2019}, authors discusses a decentralized consensus ADMM to develop a cost allocation mechanism that enable prosumers to share the cost of using common infrastructure and services for P2P trading. Meanwhile, ADMM is utilized in \cite{Hamada_AE_Apr_2019} to locally optimize reactive power compensation and active power curtailment of each inverter participating in the P2P trading for voltage control purposes. Besides, ADMM, the application of constrained optimization has also been exploited in \cite{Xu_TIE_Nov_2019} for addressing network loss issues at the physical layer.
\subsection{Blockchain}
\subsubsection{Preliminaries}Blockchain, which was first introduced in \cite{Satoshi_BitCoin_2008}, is a distributed data structure that is replicated and shared among the members of a network. With blockchain in place, applications that could previously run only through a trusted intermediary, can now operate in a decentralized fashion, without the need for a central authority, and achieve the same functionality with the same amount of certainty~\cite{Christidis_Access_May_2016}. Thus, given the properties of P2P trading, blockchain has profound applications in the future energy network. This has led to establish a number of blockchain based platforms for P2P energy trading in recent times.
\paragraph{Smart contracts}The smart contract is essentially a computerized transaction protocol that executed the terms of a contract~\cite{Szabo_Monday_Sept_1997}. By translating contractual clauses into code and embedding them into property that can enforce them, the need for trusted intermediaries between transacting parties, and the occurance of malicious and accidental exceptions can be minimized~\cite{Christidis_Access_May_2016}. Within the blockchain context, smart contracts are scripts stored on the blockchain with a unique address~\cite{LeeThomas_Nature_2019}. A smart contract is triggered by addressing a transaction to it. It then executes independently and automatically in a prescribed manner on every node in the network, according to the data that was included in the triggering transaction.
\paragraph{Elecbay}The elecbay is a software platform dedicated to the development of P2P trading within a microgrid. Each order contains the information including the time period for the energy exchange, the amount of energy to be exchanged, the price of the energy to be exchanged and the details about the seller and buyer.  After the orders are placed by peers, they are either accepted or rejected by Elecbay, based on the network constraints. After the order acceptance or rejection, each peer generates/consumes the amount of energy as promised in the accepted orders and energy is delivered through the distribution network. Further details of the platform can be found in ~\cite{Zhang_AE_June_2018}.
\paragraph{Consortium blockchain}The consortium blockchain is a specific blockchain with multiple authorized nodes to establish the distributed shared ledger with moderate cost~\cite{Kang_TII_Dec_2017}. It is established on authorized nodes to publicly audit and share transaction records without relying on a trusted third party. During P2P trading, energy transaction records among peers are uploaded to the authorized nodes after encryption. The authorized nodes run an algorithm to audit the transactions and record them into the shared ledger. This ledger is publicly accessed by participating peers and authorized nodes connected to the consortium blockchain.
\paragraph{Hyperledger}The hyperledger is an open source collaborative effort, hosted by the Linux Foundation, to advance cross-industry blockchain technologies~\cite{Hyperledger_2018}. It uses a consensus mechanism to create a transparent and non-tampering distributed ledger. According to \cite{Zhang_Hyperledger_2018}, the core module of the Hyperledger IBM runs in an open platform called Docker. When a peer within the system wants to trade, it logs into the system through the blockchain system terminal and submits the appropriate transaction. After the transaction is submitted, the transaction information is sent to a power trading unit to analyse and initiate the transaction. The information is mapped in the database in the form of key and value for user query. When mapping is complete, the power trading unit carries out the dispatching to complete the power trading.
\paragraph{Ethereum}The Ethereum, which was launched in 2015, is a programmable public blockchain with a native cryptocurrency called Ether~\cite{Ethereum_2015}. While the structure of Ethereum is very similar to that of bitcoin, a big difference is that in ethereum nodes also store the most recent state of each smart contract, in addition to all other ether transactions. For each ethereum application, the network needs to keep track of the the current information of all of these applications, including each peer's balance, all the smart contract code, and the location where it's all stored. Like bank account funds, ether tokens appear in a wallet, and can be ported to another account.

Besides the outlined classification, other modified versions of blockchain based approaches have also been used for securing energy trading in smart grid. Examples of such modified schemes can be found in \cite{Luo_TPWRS_EA_2018} and \cite{Aitzhan_TDSC_Sept_2018}.
\subsubsection{Blockchain for P2P trading in the virtual layer}In the virtual layer, existing studies have exploited a large variation of blockchain based platforms to ensure secure and transparent energy trading. In \cite{Luo_TPWRS_EA_2018}, the authors propose a blockchain based system, which is based on a parallel double-chain combined with a high frequency verification mechanism to enable a trusted and secure settlement of electricity trading transactions. To achieve a similar energy trading performance among plug-in hybrid electric vehicles within an electricity market, \cite{Kang_TII_Dec_2017} and \cite{Li_TII_Aug_2018} develop trading platforms using consortium blockchain. A multi-signature based blockchain is designed in \cite{Aitzhan_TDSC_Sept_2018} to provide transaction security in decentralized energy trading in smart grid without reliance on trusted third parties. Using smart contracts, secured P2P trading between energy storage systems and heterogeneous end-users from residential, commercial, and industry sectors are performed in \cite{Yang_TSMC_2019} and \cite{Li_AE_Aug_2019} respectively. In \cite{Wang_TSMC_Aug_2019}, the authors utilize the IBM hyperledger fabric architecture to create an operational model of crowdsourced energy systems in distribution networks considering various types of energy trading transactions and crowdsources. For secured energy trading in microgrids, Elecbay is proposed in \cite{Zhang_AE_June_2018}. Further application of blockchain based platforms for decentralized energy supply and demand management via P2P trading can be found in \cite{Devine_TSG_Mar_2019,Noor_AE_Oct_2018,Faizan_JEM_July_2019} and \cite{Zhang_AE_Sep_2018}.
\subsubsection{Blockchain for P2P trading in the physical layer}Since the physical layer is mainly responsible for accommodating the transfer of energy from sellers to buyers after the secured transactions, the focus of how the security of such transactions may impact physical layer performance is not necessary and therefore has not been reported to date. Now, while game theory, double auction, constrained optimization, and blockchain have been extensively used in the literature for designing P2P energy trading scheme, a number of new other methods are also becoming popular. Examples of such emerging techniques include graph theory~\cite{Nikolaidis_TPWRS_Early_2019}, heuristic multi-agent simulation~\cite{Zhou_AE_July_2018}, artificial intelligence~\cite{Saifuddin_Access_Apr_2019,Chen_TSG_July_2019}, and activity based models~\cite{Hermana_ITSM_Fall_2016}. 
\begin{figure*}
\centering
\includegraphics[width=\linewidth]{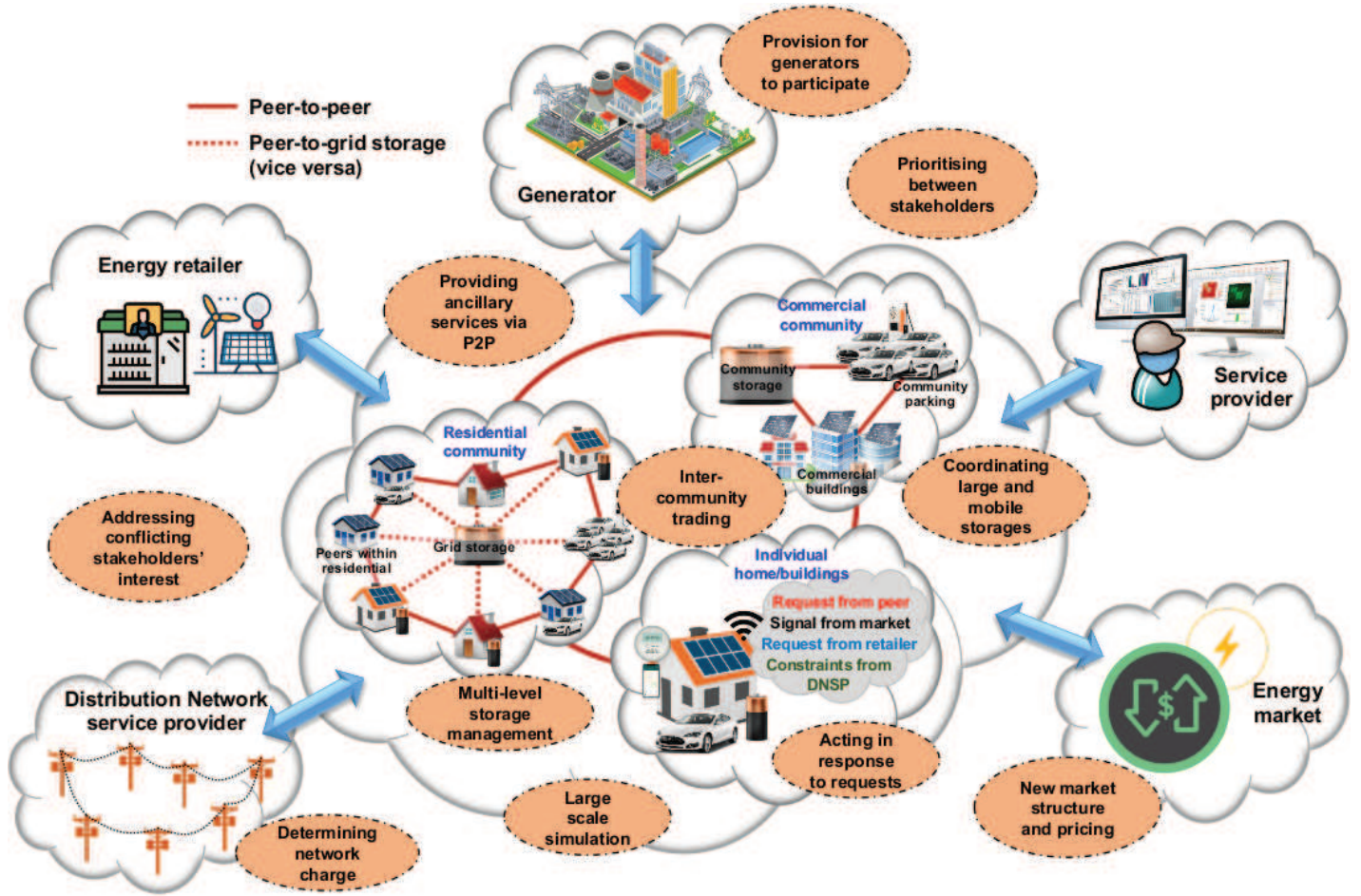}
\caption{Overview of different challenges of P2P trading for future research.}
\label{fig:Challenges}
\end{figure*}
\section{Discussion on Future Research Direction}\label{sec:FutureResearch}Note that despite extensive attention in the last couple of years, energy-management research for peer-to-peer networks is relatively new. Hence, much work is yet to be done before the integration of peer-to-peer energy trading into the current energy system. To that end, what follows is a description of a number of challenges that require further investigation as future research topics in the area of P2P trading. An overview of these challenges is also illustrated in Fig.~\ref{fig:Challenges}.
\paragraph{Network charge identification}Unlike traditional electricity systems, prosumers do not use the entire energy network for peer-to-peer trading. Therefore, the way they are currently being charged in their electricity bills needs to be researched and revised for billing them under a peer-to-peer trading paradigm.
\paragraph{Large scale network trading and simulation}Since the flow of electricity cannot be regulated, it is unlikely for a very large network that the intended receiver will receive the actual power that has been pushed to the network by the sender. Hence, the power loss due to peer-to-peer trading would be different, which necessitates further investigation. Further, P2P trading algorithms need to be simulated for large-scale realistic power system model to observe the impact of computational complexity on the conduct of trading in such a large system.
\paragraph{Benefit to the grid}Consumer-centricity of peer-to-peer trading has been well established in recent literature. However, the benefit of peer-to-peer energy trading to the distribution grid also needs to be demonstrated. Further, the grid should also have the provision to participate in P2P trading either as a generator or service provider, if necessary. This will be particularly important to pave the way for this new approach to be approved as a part of the energy system.
\paragraph{Ancillary service to the grid}Peer-to-peer trading has demonstrated potential to form coalitions between the prosumsers of the network to achieve a reliable and cost-effective supply of energy. Now, it would be an interesting extension to investigate how such coalitions can help to provide ancillary services to the grid such as with virtual power plants.
\paragraph{Multi-level storage management}With P2P trading, it is expected that a community may have different types of storage facilities, including small-scale batteries at the prosumers premises, medium-scale community storage, and large-scale grid storage. Coordination between these storage devices in an economic way and developing suitable pricing mechanisms to conduct inter-storage energy sharing would be a very complicated problem. Therefore, innovative scheduling and optimization techniques need to be developed.
\paragraph{Prioritizing stakeholders}Clearly, different stakeholders would be interested in exploiting prosumers' batteries to deliver different services to their customers and maintain network security. For example, generators may want to use them for reducing production volatility, Distribution Network Service Providers (DNSP) for demand constraint, and  retailers may want to discharge the batteries to combat energy imbalances. Nevertheless, these actions could be conflicting with one another. Hence, P2P trading schemes need to be designed such that they do not affect participants' independence and benefits.
\paragraph{Injection limit and market mechanism}At present, a cap is being set for each prosumer on the maximum amount of energy its inverter can send to the grid. This limits prosumers capacity to install larger capacity rooftop solar PV and earn more revenue. With P2P trading, prosumers can actively negotiate with one another on the amount of energy they can trade with one another and the pricing. Therefore, the injection limit needs to be flexible to extract the most benefits from such decision making processes. This necessitates the development of novel market mechanisms that will dynamically change the injection limit based on the supply and demand of energy within the network without impacting the network detrimentally.
\paragraph{Unified model}At present, research are directed either to the virtual layer or to the physical layer. However, for a successful deployment of P2P trading within the network, it is important that requirements of both layers be addressed. Hence, there is a need for a unified model, which could capture this. This is particularly possible as blockchain based information systems make all real-time system information available to both prosumers and system operator. Thus, it could be possible for the system operator to assist the participants to take decisions in the virtual layer that do not contradict with the constraints in the physical layer. Nevertheless, the extent to which a system operator can influence prosumers' behaviors needs to be carefully articulated. Otherwise, the decentricity of P2P trading could be compromised.
\paragraph{Enabling data accessibility with privacy} Of significant importance to the performance of the P2P trading, both intra-community and inter-community, is statistically useful, and accurate, energy transaction and usage data being made available across communities, for better prosumers decision making. However, this accessible data needs to also provide privacy to each prosumer. Thus provably-private transformations, of communities? energy trading data, are required to facilitate data accessibility and required sharing, while providing sufficient statistical accuracy for interrogation of the data. This remains a key research challenge, as such deployment is highly dependent on the applied scenarios for the P2P trading, and also on the desired level of utility from the privacy-preserved data.
\paragraph{Inter \& Intra-community trading}In P2P trading, a prosumer should have enough flexibility to decide whether it wants to trade with peers within its community (intra-community) or with someone external (inter-community). Market mechanisms for P2P trading should have policies and technologies ready to accommodate such flexibility.
\section{Conclusion}\label{sec:conclusion}In this review article, an overview of existing research in peer-to-peer energy trading has been provided. As such, first, the background of peer-to-peer trading in energy networks has been discussed with specific emphasize on features of peer-to-peer networks, market structure for peer-to-peer trading, and opportunities and challenges. Second, a systematic classification of peer-to-peer energy has been proposed based on the relevant challenges in both virtual and physical layers that have been addressed by the state-of-the-art research papers. Third, core technical approaches that have been extensively used in the literature have been identified and summarized. In addition, an overview of the application of these identified approaches for peer-to-peer trading in virtual and physical layers has been provided. Finally, this paper have discussed a number of interesting topics for researchers to work on in the future.
%\def\baselinestretch{.94}
%\bibliographystyle{IEEEtran}
%\bibliography{NatureEnergy}

\end{document}